\documentclass[5p,twocolumn,times,number]{elsarticle}

\usepackage{graphicx}
\usepackage{amsmath}   

\begin{document}

\begin{frontmatter}

\title{ Energy and time resolution for a LYSO matrix prototype of the Mu2e 
  experiment }

\author[a]{N.~Atanov}
\author[a]{V.~Baranov}
\author[b]{F.~Colao}
\author[b]{M.~Cordelli}
\author[b]{G.~Corradi}
\author[b]{E.~Dan\'e}
\author[a]{Yu.I.~Davydov}
\author[c]{K.~Flood}
\author[b]{S.~Giovannella\corref{cor}}
\ead{simona.giovannella@lnf.infn.it}
\author[a]{V.~Glagolev}
\author[b]{F.~Happacher}
\author[c]{D.G.~Hitlin}
\author[b,d]{M.~Martini}
\author[b]{S.~Miscetti}
\author[c]{T.~Miyashita}
\author[e,f]{L.~Morescalchi}
\author[g]{P.~Ott}
\author[e,h]{G.~Pezzullo}
\author[b]{A.~Saputi}
\author[b]{I.~Sarra}
\author[b]{S.R.~Soleti\corref{cor}}
\ead{roberto.soleti@lnf.infn.it}
\author[i]{G.~Tassielli}
\author[a]{V.~Tereshchenko}
\author[g]{A.~Thomas}

\cortext[cor]{Corresponding author}

\address[a]{Joint Institute for Nuclear Research, Dubna, Russia}
\address[b]{Laboratori Nazionali di Frascati dell'INFN, Frascati, Italy}
\address[c]{California Institute of Technology, Pasadena, United States}
\address[d]{Universit\`a ``Guglielmo Marconi'', Roma, Italy}
\address[e]{INFN Sezione di Pisa, Pisa, Italy}
\address[f]{Dipartimento di Fisica dell'Universit\`a di Siena, Siena, Italy}
\address[g]{Institut f\"ur Kernphysik, University of Mainz, Mainz, Germany}
\address[h]{Dipartimento di Fisica dell'Universit\`a di Pisa, Pisa, Italy}
\address[i]{INFN Sezione di Lecce, Lecce, Italy}


\begin{abstract}
We have measured the performances of a LYSO crystal matrix prototype
tested with electron and photon beams in the energy range 60$-$450 MeV.
This study has been carried out to determine the achievable energy and
time resolutions for the calorimeter of the Mu2e experiment. 
\end{abstract}



\end{frontmatter}

\section{Introduction}

A $5\times 5$ matrix prototype, built with ($30\times 30\times 200$) 
mm$^3$ LYSO scintillating crystals from SICCAS, was assembled in Frascati 
to study the performances of such a calorimeter for the Mu2e experiment 
\cite{NimCaloMu2e}.
Global transverse and longitudinal dimensions provide a coverage of
$\rm \sim 3.8\ R_M$ and $\rm 11.2\ X_0$, respectively.
Although the baseline crystal is now BaF$_2$ \cite{TDR}, tests with 
electron and photon beams have been carried out to measure the 
calorimeter performances.
Before the assembly, the crystals have been extensively tested using a
$^{22}$Na source and a spectrophotometer. All of them showed good light 
yield, longitudinal response uniformity and transmittance.
Each crystal has been wrapped with 60 $\mu$m thick Enhanced Specular 
Reflector (ESR) from 3M and then optically connected to a Hamamatsu 
S8664 large area avalanche photodiode (APD) using Saint-Gobain BC-630 
grease. Amplification and bias voltage regulation was provided by custom 
made front end electronic boards.

The LYSO matrix has been tested with tagged photon beams in the energy 
range 60$-$190 MeV at MAMI \cite{MAMI} (Mainz, Germany) and with 
80$-$450 MeV electron beams at BTF \cite {BTF} (Frascati, Italy). Here, 
the trigger was provided by two orthogonal ($6\times 10\times 50$) mm$^3$ 
plastic scintillation counters read out by ($3\times 3$) mm$^2$ silicon 
photomultipliers.
Data were acquired with CAEN V1720 waveform digitizer, 250 Msps, 12 bit
resolution and 0$-$2 V dynamic range. APDs were illuminated, through 250 
$\mu$m diameter fused silica optical fibers, by a green laser 
($\lambda = 530$ nm), whose pulse was synchronized with an external 
trigger at a frequency of $\sim 1$ Hz.
Equalization of matrix channels at 10\% level was obtained using minimum 
ionizing particles (MIPs) crossing vertically the detector. Calibration 
 of cell response was done directly with beams (450 MeV electrons at BTF, 
92.5 MeV photons at MAMI) firing on each cell center.

\section{Test beam results}

Beam events have been selected with a cut on the waveform time distribution,
which retains the range corresponding to beam particles hitting the matrix.
The total charge for each crystal is defined as the sum of the waveform
spectrum in a selected time range, 400 ns wide, around the peak. Baseline
is evaluated in a same width region, far from the peak. Multiple scattering 
events has been reduced by cutting on the distance between the energy 
weighted centroid and the impact point in the calorimeter, which is kept 
below 0.5 cm. For BTF beams, the particle multiplicity, $\rm\mu_p$, is 
greater than one and it can be tuned by adjusting beam intensity and 
collimators. In our test it was set to $\rm\mu_p\sim 0.8$.
Peaks due to one, two and three particle events are clearly visible and 
well separated in the energy spectra. In this analysis, single and double
particle events have been selected with a cut in the total charge of both 
scintillator counters and matrix.

\begin{figure}[!t]
\vspace{-5mm}
\centering
\includegraphics[width=0.99\linewidth]{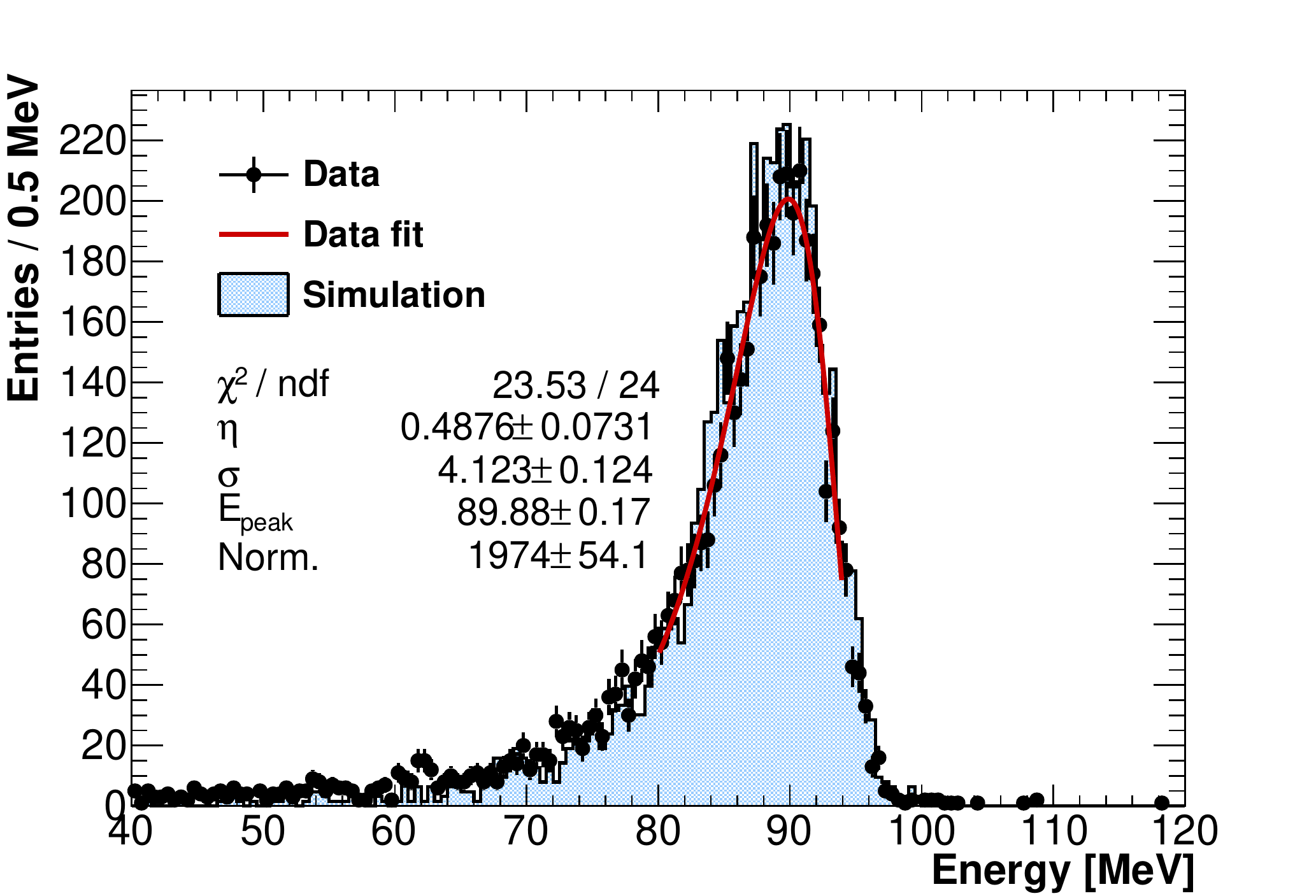}
\caption{Energy distribution for 90 MeV photons (dots) compared with 
GEANT4 simulation (filled histogram). MC spectrum includes 2 mm beam 
spread and an additional 2.6\% Gaussian smearing accounting for 
miscalibration, non uniformity and non linearity. Energy resolution 
is obtained from a fit with a Lognormal distribution (solid line).}
\label{Fig:EneSpectrum}
\end{figure}

The energy scale has been set, after the offline equalization, by 
comparing the total reconstructed charge in the matrix with the expected 
energy deposited in the entire matrix, $E_{\rm dep}$, as estimated by a 
GEANT4 simulation. Besides the beam spatial spread, an additional 
constant 2.6\% Gaussian smearing is needed in the simulation to reproduce 
real data, accounting for miscalibration, non uniformity and non 
linearity (Fig.~\ref{Fig:EneSpectrum}).

\begin{figure}[!t]
\vspace{-5mm}
\centering
\includegraphics[width=0.99\linewidth]{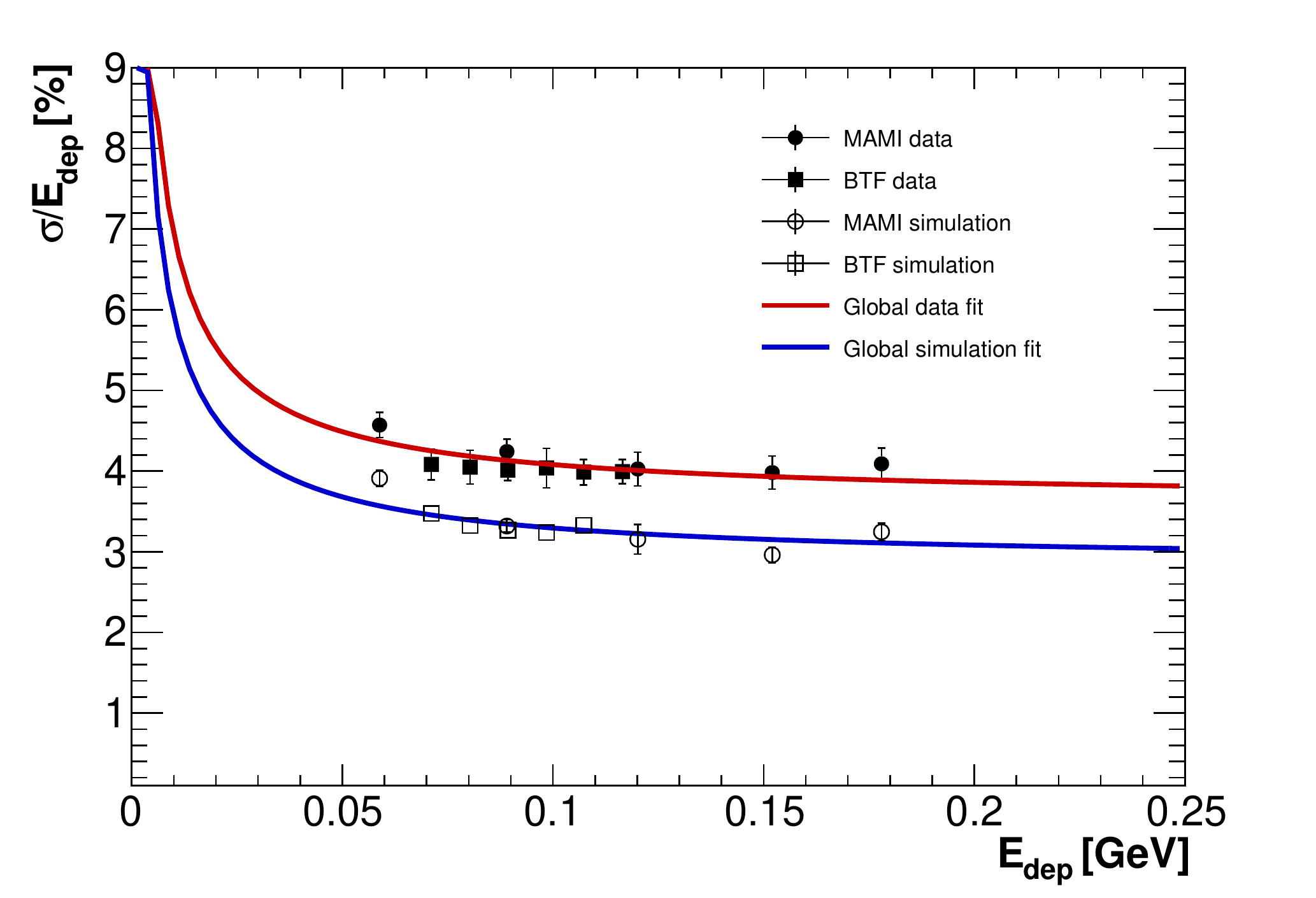}
\caption{Energy resolution as a function of the deposited energy for 
$\gamma$'s (dots) and $e^-$ (full squares). Corresponding MC 
expectations are reported in circles and open squares, respectively.
Simulation points are obtained without the 2.6\% Gaussian smearing 
needed to describe real data.}
\label{Fig:EneReso}
\end{figure}

The energy resolution has been obtained from a fit with a Lognormal 
distribution to the energy spectra. While at MAMI the beam energy 
spread is negligible, at BTF it is of the order of 5\% in our energy 
range. Therefore, for each energy, the intrinsic energy spread of the 
beam, $\sigma_{\rm b}(E)$, has been obtained by comparing the energy 
resolution of one-particle and two-particle events. After subtracting 
$\sigma_{\rm b}(E)$, the energy resolution is reported as a function of 
the deposited energy in Fig.~\ref{Fig:EneReso} both for electron and 
photon beams. In the same figure, the corresponding simulated events 
without including the Gaussian smearing are also reported.
A fit is then  performed with the formula:
\begin{equation}
\frac{\sigma_{E}}{E_{\rm dep}} = \frac{a}{\sqrt{E_{\rm dep}(\rm GeV)}} \oplus\, b.
\end{equation}
The extracted parameters are: $a=(0.6\pm0.1)\%$, $b=(3.6\pm0.2)\%$.
The same parametrization on Monte Carlo (MC) events
provides $a=(0.52\pm0.04)\%$, $b=(2.86\pm0.09)\%$.

\begin{figure}[!t]
\centering
\includegraphics[width=0.99\linewidth]{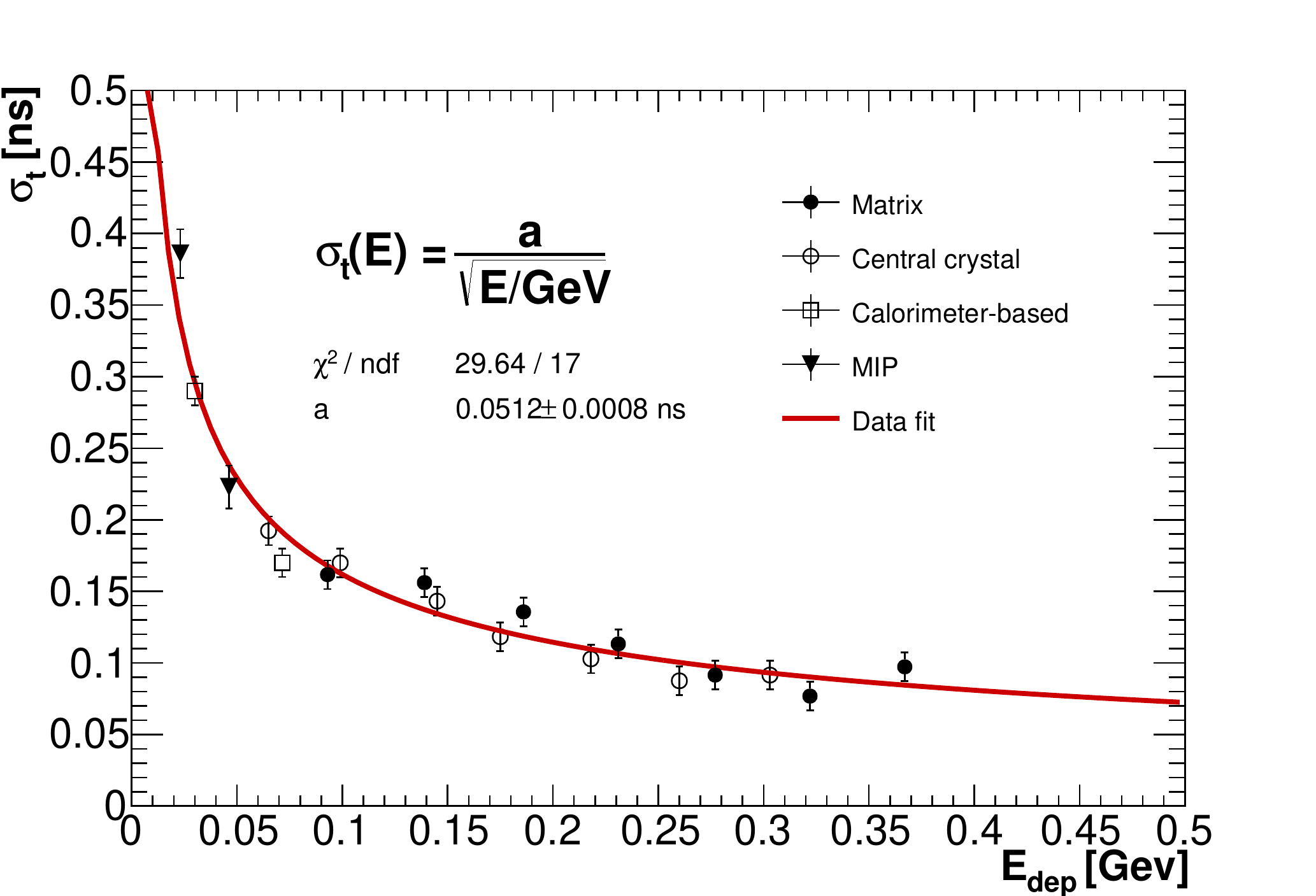}
\caption{Time resolution for $e^-$ as a function of the deposited 
energy. The jitter due to the trigger has been subtracted. The 
$\sigma_T$ value has been obtained with the whole matrix (dots),
the central crystal (circles) and the difference of two crystals 
when the beams hit in the middle of the two (open squares). The
low energy region has been exploited with minimum ionizing particles
(triangles).}
\label{Fig:TimeReso}
\end{figure}

The time of the event has been extracted from a fit to the waveform shape 
of the digitizer, performed with a Landau function. A residual time walk 
is observed, so that slewing corrections are applied. 
Time resolution has been measured with electrons using both the central 
crystal and the energy-weighted time of the whole matrix. In both cases, 
the time of the external trigger, defined as the semi-sum of the time for 
scintillation counters, has been subtracted event by event.
The time jitter of the trigger has been measured from a Gaussian fit to 
the time difference of the two scintillation counters: $(145 \pm 2)$ ps.
After subtracting the trigger jitter, the calorimeter time resolution 
as a function of the deposited energy is reported in Fig.~\ref{Fig:TimeReso}.
To exploit the low energy region, the time resolution has been also
evaluated with minimum ionizing particles and with special runs where 
the beam hits in the middle of two crystals, using their time difference.
Data have been fit taking into account the stochastic contribution only,
and are well parametrized by the scaling law:
\begin{equation}
\sigma_T = (51\pm 1)\ {\rm ps} / \sqrt{E_{\rm dep}(\rm GeV)}.
\end{equation}


The measured performances well satisfy the requirements of the Mu2e
calorimeter, that are set to $\sigma_E / E = 5\%$ and $\sigma_T < 500$ ps
for 100 MeV electrons.




\end{document}